# Positive Feedback Regulation Results in Spatial Clustering and Fast Spreading of Active Signaling Molecules on a Cell Membrane

Jayajit Das[1#], Mehran Kardar[4], Arup K. Chakraborty[1-3,*]

Departments of [1]Chemical Engineering, [2]Chemistry, [3]Biological Engineering, & [4]Physics, Massachusetts Institute of Technology, Cambridge, MA 02139.

* address correspondence to arupc@mit.edu

## ABSTRACT

Positive feedback regulation is ubiquitous in cell signaling networks, often leading to binary outcomes in response to graded stimuli. However, the role of such feedbacks in clustering, and in spatial spreading of activated molecules, has come to be appreciated only recently. We focus on the latter, using a simple model developed in the context of Ras activation with competing negative and positive feedback mechanisms. We find that positive feedback, in the presence of slow diffusion, results in clustering of activated molecules on the plasma membrane, and rapid spatial spreading as the front of the cluster propagates with a constant velocity (dependent on the feedback strength). The advancing fronts of the clusters of the activated species are rough, with scaling consistent with the Kardar-Parisi-Zhang (KPZ) equation in one dimension. Our minimal model is general enough to describe signal transduction in a wide variety of biological networks where activity in the membrane-proximal region is subject to feedback regulation.

# Present Address: Battelle Center for Mathematical Medicine, The Research Institute at the Nationwide Children's Hospital, Departments of Pediatrics, Physics and Biophysics Graduate Program, Ohio State University, 700 Children's Drive, Columbus, OH 43205

# INTRODUCTION

Cell signaling networks often generate binary (on or off) responses in presence of a diverse set of stimuli in the local environment. A common element present in many of these networks is a positive feedback loop [1,2], which can give rise to discrete decisions [1,3,4]. A positive feedback loop can arise when an activated signaling molecule creates a mediatory molecule that in turn enhances the activation of the signaling molecule. When the timescales of the biochemical reactions involved are slower, or of the same order, as that of the diffusion of the molecules participating in the reactions, nonlinearities associated with positive feedback could couple with diffusion to result in spatial clustering, and in spreading of activated molecules at a rate much faster than diffusion. Various models describing growth of advantageous alleles in the area of population biology[5], kinetics of reaction fronts in autocatalytic systems[6,7], spreading of bacterial colonies[8], and pattern formation during embryonic development[9], have incorporated the coupling of nonlinearities with diffusion in their dynamics. It was also found that auto catalytic reactions in the presence microscopic discreteness could give rise to spatial clustering of active species[5,6]. These models are naturally constructed to explore the behavior of systems at the length and time scales relevant to the population, organism, or cellular level. The role of positive feedback in spatial dynamics of sub-cellular processes, relevant for signal transduction in cell signaling networks, has begun to be appreciated only very recently[10,11]. In this paper, we study the spatial-temporal evolution of cell signaling dynamics subject to feedback regulation when diffusive processes occur on time scales similar to the signaling reactions. The diffusion of molecules is usually much slower (about ~100 times [12]) in the plasma membrane compared to the cytosol, thus the effects we described are most relevant for the molecules in the plasma membrane participating in membrane proximal cell signaling.

**MODEL**

As a prototype of signaling events on a two-dimensional cell membrane that involves positive feedback regulation, we consider the activation of the membrane associated Ras family of proteins. Ras can be activated by a Guanine exchange factor protein, Son of Sevenless (SOS). Specifically SOS catalytically converts GDP bound Ras to its GTP bound activated form. It was discovered recently [13,14] that catalysis of Ras-GDP to Ras-GTP aided by SOS becomes even faster (~75 fold) when a membrane associated SOS molecule is bound to Ras-GTP at an allosteric site. This mechanism introduces positive feedback regulation of Ras activation[15]. Activated Ras has been observed to form clusters, and the diffusion coefficient in these clusters can be small.

We study the following simplified set of three reactions that aim to mimic positive feedback, as in Ras activation:

$$Z + Y \xrightarrow{k_1} X + Y, \quad Z + X + Y \xrightarrow{k_2} 2X + Y, \quad Y \xrightarrow{k_3} \varnothing \ . \tag{1}$$

In the above reactions, the Y species can be thought of as representing SOS, while Z and X are analogous to Ras-GDP and Ras-GTP respectively. The first two reactions in Eq.(1) correspond to activation of Ras without and with positive feedback, respectively. The last reaction describes detachment of SOS from the plasma membrane upon phosphorylation by activated Erk, a transcription factor that gets activated as a result of Ras activation. The reaction scheme is quite general and can be applied to a large variety of competing positive and negative feedbacks found in cell signaling. In such cases, when the diffusion time scale is much faster than the reaction time scales in Eq.(1), stochastic fluctuations in the feedback reactions can give rise to binary outcomes (activation vs. de-activation) when the number of molecules is small[3]. Here we show that the opposite limit, when diffusion time scales are slower than the reaction times, give rises to spatial clustering, and enables signal propagation by advancing cluster at a much faster rate than diffusion. We also find the growing fronts become rough, in a manner consistent with the Kardar-Parisi-Zhang (KPZ) equation[24,25].

**METHOD**

We construct a lattice model in one (d=1) and two (d=2) dimensions to study the model described above using a kinetic Monte Carlo (MC) simulation. The system with d=2 describes signaling on a cell membrane, and the one dimensional system may be relevant if the molecules are constrained to move in narrow channels. For ease of presentation, the MC scheme in described for d=1 (Fig.1), but can be easily generalized to higher dimensions. At t=0, a lattice point (site *i*) can be unoccupied or occupied by a particle of X, Y or Z species. A site cannot be occupied by more than one particle because of their hardcore repulsion. We have chosen initial concentrations of Z and Y particles as 0.469 and 0.03, respectively. All the simulations are performed with an initial homogeneous distribution of Z and Y particles. At every MC trial a lattice site is picked at random, and if its occupancy ($n_i$) is non-zero, we attempt the following moves with equal probability by calling a random number $r \in (0,1]$ from a uniform distribution. (i) If $0 < r \leq 1/4$, a nearest neighbor point ($i+1$ or $i-1$ with probability 1/2) is selected, if that site is empty ($n_{i+1\,or\,i-1} = 0$), then the diffusion move (exchange of particle occupancy between $i$ and, $i+1$ or $i-1$) is accepted with a probability $p_D = \Delta t D^{MC}$, where, $\Delta t = 1/(k_1^{MC} + k_2^{MC} + k_3^{MC} + D^{MC})$, when all the species have the same diffusion constant, $D^{MC}$. The parameters, $k_1^{MC}$, $k_2^{MC}$, $k_3^{MC}$, and $D^{MC}$, used in the MC trials are related to the physical rate constants (in d-dimensions) by, $k_1 = 2k_1^{MC} dl_0^d$, $k_2 = 6k_2^{MC} dl_0^{2d}$, $k_3 = k_3^{MC}$, and $D = D^{MC} l_0^2$ respectively, where $l_0$ denotes the lattice spacing in the simulation and the time scale is set by $\Delta t$. (The details of this derivation are left to the appendix.) We also study the case when the Y species has a different diffusion constant, $D_Y^{MC}$, in which case $\Delta t = 1/(k_1^{MC} + k_2^{MC} + k_3^{MC} + D_{max}^{MC})$, where, $D_{max}^{MC}$ takes the value of the larger diffusion constant between $D^{MC}$ and $D_Y^{MC}$. The diffusion moves for X or Z species and Y species are chosen with a probability, $p_D = \Delta t D^{MC}$, and $p_{D_Y} = \Delta t D_Y^{MC}$ respectively. (ii) If $1/4 < r \leq 1/2$, a nearest neighbor point ($i+1$ or $i-1$) is selected, if that site is occupied ($n_{i+1\,or\,i-1} = 1$), and $i$ and $i+1$ (or $i-1$) sites have a pair of Z and Y particles, then the first reaction in Eq.(1) is executed with a probability $p_{r_1} = k_1^{MC} \Delta t$. (iii) If $1/2 < r \leq 3/4$, we attempt the second reaction in Eq.(1). Any of the pairs, $\{i+1, i-1\}$, $\{i+1, i+2\}$, and $\{i-1, i-2\}$ is chosen with equal probability, and, if all the three sites (site $i$ and the chosen pair of sites) are occupied with X,Y and Z particles, the reaction is executed with a probability $p_{r_2} = k_2^{MC} \Delta t$. (iv) If, $3/4 < r \leq 1$, and if the site $i$ is occupied by a particle of Y species, we attempt the last reaction in (1) with probability $p_{r_3} = k_3^{MC} \Delta t$.

While the model we study is quite general, we have studied it using parameters relevant to Ras activation. The production of Ras-GTP from Ras-GDP through SOS can be represented by the reaction scheme in Eq.(1) as shown in the appendix, where, Z, X and Y particles represent Ras-GDP, Ras-GTP and SOS molecules, as indicated before. The model parameters obtained from the rate constants measured in vitro are, $k_1 = 3.29 \times 10^5 (\mu m)^2 (molecules)^{-1} s^{-1}$ and $k_2 = 4.62 \times 10^3 (\mu m)^4 (molecules)^{-2} s^{-1}$. The diffusion constant of Ras molecules varies from 0.01 $(\mu m)^2/s$ to 1 $(\mu m)^2/s$ [16]. Therefore, we have chosen units such that, $l_0 = 1$ (one lattice spacing) and $D^{MC} = 10$ in the simulations correspond to $l_0 = 0.1 \mu m$ and $D = 0.1 (\mu m)^2 s^{-1}$ in actuality. This choice also implies that, $k_1^{MC} = 0.001$, $k_2^{MC} = 10.0$ and a concentration $\rho_Z = 0.469$ of Z particles in simulations correspond to $k_1 = 4.0 \times 10^5 (\mu m)^2 (molecules)^{-1} s^{-1}$, $k_2 = 12.0 \times 10^5 (\mu m)^4 (molecules)^{-2} s^{-1}$, and a Ras concentration of 47 molecules/$(\mu m)^2$ respectively in experiments. These estimates are within realistic ranges of the parameters for cellular systems, and we use them for most of the simulations presented in the main text. More details of this particular choice of parameters are shown in the appendix. We found the qualitative features of our results do not change as the above parameters are varied over at least a factor of 10. As long as feedback is sufficiently strong, the phenomena we describe are robustly reproduced.

**RESULTS:**

**Domain growth in two dimensions**

Spatial dimension plays a crucial role in controlling the dynamics of the system. In d=1, the particles get locked in arrested states indefinitely, even when the Y particles are spared of the annihilation reaction ($k_3^{MC} = 0$). Though this scenario is not relevant for the molecules interacting in a cell's plasma membrane, such a situation may arise if signaling molecules are constrained to move in narrow channels. The physical reason underlying this behavior in d=1 is the following: When a Y particle is flanked by two neighboring X species (Fig. 1c (iv)), the Y particle cannot participate in any reaction and is rendered "inactive" indefinitely because of the topology of one dimension and the hardcore interactions between the particles. Thus, when all the Y particles become "inactive", all the reactions in the system come to a stop leading the system into the arrested states.

We focus on d=2, where there are no corresponding geometrical constraints for inevitable arrested states. Instead, the system displays nucleation and domain growth: initially X particles are created from Z particles through the first reaction in Eq.(1). Clusters of X particles may seed a "critical nucleus" that grows with time because of the positive feedback. The nucleation and growth of domains of the X particles is shown in Fig.2a. In the absence of positive feedback ($k_2^{MC} = 0$), X particles are created, but there is no nucleation or domain growth in the system (Fig. 2b).

The concept of a critical nucleus size can be formulated in the following way. If a cluster of X particles of size $l$ is created, it can increase in size because the X particles at its boundary convert the nearby Z particles (in presence of Y particles, which are homogenously distributed throughout the region) into X particles by the positive feedback reaction. However, it may also shrink in size as the X particles can also diffuse away from the cluster. The time scale for the removal of a particle by diffusion from a region of size is $\tau_D \sim l^2/D$, while the time scale associated with creation of X particles from Z particles through positive feedback reaction is $\tau_{feedback} \sim 1/(k_2 \bar{\rho}_Y \bar{\rho}_Z)$, where $\bar{\rho}_Y$ and $\bar{\rho}_Z$ are the average densities of the Y and Z particles respectively. Thus, if $\tau_D \gg \tau_{feedback}$, the X particles in the cluster will diffuse away before they can increase the size of the cluster by positive feedback. This suggests that clusters of size,

$$l > l_c = (D/k_2 \bar{\rho}_Y \bar{\rho}_Z)^{1/2}, \qquad (2)$$

will grow due to positive feedback regulation. The decay of the Y particles limits the time available for growth of the domains of X particles. This is because; once all the Y particles are annihilated, no new X particles can be created. Thus a larger value of the constant $k_3$ results in a smaller size of the largest X domains (Fig.S1 in EPAPS). Therefore, to characterize the scaling properties of domain growth (e.g., size of the critical nucleus, or growth law), we set $k_3^{MC} = 0$. This is a choice of convenience, and does not alter the general principles emerging from our study.

In order to quantify the size of the critical nucleus, and the growing domains, we calculate the dynamic structure factor defined as, $S(q,t) = \langle \rho_X(\vec{q},t)\rho_X(-\vec{q},t) \rangle$, where, $\rho_X(\vec{q},t)$ is the Fourier transform of the density $\rho_X(\vec{r},t)$ of X particles; and $\langle \cdots \rangle$ denotes averaging over initial configurations. We employ methods similar to those used to deduce the size of the critical nucleus from the structure factor at a first order equilibrium phase transition[17]. The key concept behind this method as follows: Any domain of size, $l \sim 1/q > 1/q_c$, where $q_c$ is the wave-vector corresponding to the size, $l_c$, of the critical nucleus, will grow with time while domains of size $l < l_c$ will dissolve in the unstable phase. Therefore, if $S(q,t)$ is graphed at different times, all the curves should merge at the wave-vector, $q = q_c$. We apply the same methodology to characterize nucleation and domain growth in our system, although, unlike the previous case[17], we follow a transition from a non-equilibrium unstable steady state to a stable steady state.

Our characterization of nucleation and growth of domains from $S(q,t)$ are summarized in Fig. 3. As depicted in Fig.3a, $S(q,t)$ increases rapidly for small $q$, but gradually (roughly by a constant amount with time) at large $q$. The difference in the evolution of $S(q,t)$ for $q > q_c$, between our system and one undergoing an equilibrium first order phase transition (such as an Ising model), comes from the ever increasing number of X particles. In the equilibrium first order transition, domains smaller than the critical size dissolve and disappear into the unstable phase. In our case these domains disintegrate into even smaller clusters of the stable state (X particles), resulting in a constant vertical shift, $\Delta(t)$, in $S(q,t)$ as time increases. If we adjust each curve in $S(q,t)$ for $\Delta(t)$ at different times, we see a pattern similar to a system undergoing a first order phase transition, with the curves merging at a single wave-vector, $q = q_c$ (Fig.3b). The size of the critical nucleus increases as the diffusion constant, $D^{MC}$, of the particles is increased (Fig.3c). The size of the critical nucleus, $l_c = 2\pi/q_c$, changes with the diffusion constant of the particles (Fig. 4d), but the increase in $l_c$ does not quite follow the dimensional analysis leading to Eq.(2). We have also calculated variations of $q_c$ as the densities of Y and Z particles are changed (Fig. S2 in the EPAPS), again observing deviations from the behavior predicted by Eq. (2). Such a departure from simple scaling may arise due to a combination of non-linearity and stochastic fluctuations[18], and deserves further analysis. Also note, that in the above calculation of the size of the critical nucleus, it has been assumed that, as in the Ising system[19], the nucleation process is described by a single reaction co-ordinate, the size of the nucleus. It will be interesting to check the validity of this assumption using transition path sampling[20] like techniques that have been applied in biochemical reaction networks[21].

For $k_3^{MC} = 0$ in two dimensions, the number $N_x(t)$, of X particles monotonically increase in time. As depicted in Fig.4 for various values of the diffusion constant $D^{MC}$, the fraction of the X particles $f_X(t) = N_x(t)/M$, where $M$ is the total number of Z particles at $t = 0$, increases with time in a sigmoid manner, and saturates to unity. While such a behavior also follows from a mean-field treatment, the dependence on the diffusion constant indicates the importance of the nucleation and growth of clusters. We attempt to quantify the growth of clusters from the data in Fig.4 in the following way. Let us assume, for simplicity, that all the X particles at a time $t$ are created from the growth of a fixed number $n$ of circular critical nuclei that arise at time $t = t_0$. We can then relate increase in $f_X(t)$ to the rate of the growth of the circular domains by[22]

$$\frac{df_X(t)}{dt} = kn \frac{dA(t)}{dt}(1 - f_X(t)) \quad , \tag{3}$$

where $A(t) = \pi R^2(t)$, $R(t)$ is the radius of each island of X particles, $k$ is the proportionality constant. If we assume domains of size $R_0$ are nucleated at $t = t_0$, such that $f_X(t_0) = \pi n R_0^2 / M$ and $R(t_0) = R_0$, we can solve Eq.(3) to get

$$f_X(t) = 1 - (1 - f_X(t_0)) \exp[-kn(A(t) - A(t_0))]. \tag{4}$$

Fits to the above equation are also depicted in Fig.4. Note that approximating the complicated spatial arrangements in Fig.2a by a fixed number of growing domains is neither correct at short times (before nuclei are formed) or at long times (when the clusters merge). Thus the model can only be regarded as an approximation of intermediate times, and indeed the fits have been made so as to best match the rising portion of the numerical curves in Fig.4. The variations in time in Eq.(4) are encoded in the increase of the area $A(t)$, and hence the radius $R(t)$. We now assume a form $(R(t) - R_0) \propto (t - t_0)^a$, and extract $a$ from the fits to the data in Fig.4. We find that a linear growth of the domains, i.e., $R(t) \propto t$, can fit the data to some extent. The many sources of deviation from the form predicted by Eq.(4) include the non-circular shape of the domains (rough walls), inhomogeneous nucleation of the domains of the X particles, and merging of domains. In the next section, we address the roughness of a single interface between the Z and X particles due to positive feedback and diffusion alone.

**Fluctuations and interface motion**:

In the last section we observed that positive feedback leads to formation and growth of domains. Here, we attempt to quantify the motion and shape of the advancing interface of the stable phase (region filled with X) into the unstable phase (region filled with Z), focusing on the positive feedback reaction by itself. Specifically, we start with an initial configuration in a simulation box of size $L_x \times L_y$, where, one quarter of the box ($0 \leq x \leq L_x/4$, and $0 \leq y \leq L_y$) is occupied only by X particles and the other half of the box ($L_x/4 < x \leq L_x$, and $0 \leq y \leq L_y$) by Z particles. The Y particles are distributed homogenously in both the compartments. The system then evolves according to the reactions in Eq.(1), with $k_1=0$ (to prevent nucleation of additional domains of X particles in the compartment filled with Z particles) and $k_3=0$. Thus, the spreading of the X particles to the space initially occupied by Z particles occurs only due to positive feedback and diffusion. We study both the cases, when (A) the diffusion constants of the particles are equal, $D_Y^{MC} = D_X^{MC} = D_Z^{MC} = D^{MC}$; and (B) with $D_X^{MC} = D_Z^{MC} = D^{MC}$ and $D_Y^{MC}$ different from $D^{MC}$. Periodic boundary conditions are imposed along the $y$ direction, and reflective boundary conditions are applied at the $x$ boundaries.

We limit the maximum time reached in our simulation such that the interface between the X and Z regions does not cross the boundary at $x = L_x$. In the mean-field approximation, the dynamics of the system is described in terms of the densities of X, Y and Z particles by the set of reaction diffusion equations:

$$\frac{\partial \rho_X}{\partial t} = D \nabla^2 \rho_X + k_2 \rho_Y \rho_X (\rho_0 - \rho_X) \tag{5a}$$

$$\frac{\partial \rho_Y}{\partial t} = D_Y \nabla^2 \rho_Y \quad , \tag{5b}$$

where, $\rho_X$ and $\rho_Y$ denote the densities of the X and Y particles respectively. The total number of the Z and X particles is conserved, and on average $\rho_X + \rho_Z = \rho_0$, where, $\rho_Z$ is the density of the Z particles. The constants $D$, $D_Y$ and $k_2$ are related to the parameters used in the MC simulation as, $D = D^{MC} l_0^2$, $D_Y = D_Y^{MC} l_0^2$ and $k_2 = 6 k_2^{MC} d l_0^{2d}$ respectively, as shown in the appendix.

Equation (5a) is closely related to the well-known Fisher equation, which is a generic model for dynamic processes involving advance of a stable phase into an unstable region, and has been applied to a wide range of phenomena in biology, chemistry, and physics. It is a phenomenological description of the time evolution of a density field, $\rho(r,t)$, $r \in R^d$, where, a stable phase, $\rho = \rho_s$, propagates into the unstable phase, $\rho = 0$, following

$$\frac{\partial \rho}{\partial t} = D \nabla^2 \rho + k \rho (\rho_s - \rho) \quad . \tag{6}$$

While the Fisher equation cannot be solved exactly, stability analysis shows that a sharp interface, $\rho(r,t)=0$ for $r_{\parallel} \leq 0$, $r \in (r_{\parallel}, r_{\perp})$, $r_{\parallel} \in R; r_{\perp} \in R^{d-1}$ and $\rho = \rho_s$ for $r_{\parallel} > 0$, at t=0, will move into the unstable phase ($\rho = 0$) in the transverse direction, $r_{\parallel}$, with a velocity, $v \geq v_{min} = 2\sqrt{Dk\rho_s}$, as the width of the interface broadens to a size of $\sqrt{D/(k\rho_s)}$ at late times. Comparison of Eq.(6) with Eq.(5a) suggests that when Y particles are homogeneously distributed in the system, i.e., for $\rho_Y(r,t) = \bar{\rho}_Y = const.$, the interface between X and Z rich regions should propagate into the Z compartment with a velocity $v \geq v_{min} = 2\sqrt{\rho_Y \rho_0 D k_2}$. However, the stochastic fluctuations originating both from the reactions and the diffusion of the particles should modify this conclusion, and we study the effect of fluctuations by examining the velocity and the width of the interface from our kinetic MC simulations. We define the leading edge of the advancing domain by a set of height variables $(\{h_j\}, j=1...L_y)$. Each $h_j$ denotes the $x$ coordinate of the rightmost X particle, with $y$ coordinate equal to $j$. We then compute the average position ($\bar{h}(t)$) and the average width ($w(t)$) of the front from the height variables as

$$\bar{h}(t) = \left\langle \frac{1}{L_y} \sum_{j=1}^{L_y} h_j \right\rangle \quad , \quad \text{and}$$

$$w^2(t) = \left\langle \frac{1}{L_y} \sum_{j=1}^{L_y} (h_j - \bar{h}(t))^2 \right\rangle ,$$

where, $\langle ... \rangle$ denotes the average over a set of runs (initial configurations).

In the simulations, the front moves with a constant velocity after a short transient time. This transient period is larger for smaller values of $L_y$. The graph of $\bar{h}(t)$ with time in Fig.5a shows that the interface advances with a constant velocity, irrespective of the value of $L_y$. The velocity of the interface calculated from the slope of Fig.5a is plotted at several values of $D_Y$ in Fig.5b. We find that the front moves with velocities smaller than the minimum velocity predicted by the mean-field analysis of the reaction diffusion equation. Such a departure from the mean-field description has been also observed in other reaction diffusion systems[7,23]. Furthermore, while the mean-field analysis predicts that the velocity of the front does not depend on $D_Y$, our simulation shows (Fig.5b) that the front velocity increases with $D_Y$, and ultimately saturates to a constant value. Since the reaction front propagates in the unstable phase because of the reactions occurring at the leading edge, where the particle concentration is very low, both molecular discreteness[24] and microscopic fluctuations[7,25] arising from reactions and diffusion of the particles, play important roles in determining the velocity of the wave front. However, the actual front velocity depends on these effects, as well as on the details of the implementation of microscopic dynamics in a lattice model[7], in a complicated manner. Our results demonstrate that these factors clearly play an important role in modulating the reaction kinetics of the system.

We can try to understand the effect of mobility of the Y particles on the propagation of the interface heuristically as follows[26]. A pair of X and Z particles (denoted as, (XZ)), in presence of a Y particle, is transformed into a pair of X particles (denoted as (XX)). This process can be represented by the set of reactions,

$$XZ + Y \xleftrightarrow{(d^+, d^-)} XZY \xrightarrow{k_2} XXY, \qquad (7)$$

where, $d^+ = 2\pi(D_{XZ} + D_Y)$, $d^- = 2(D_{XZ} + D_Y)/l_0^2$, $D_{XZ}$ is the diffusion constant of the XZ particle pair, and $l_0$ is the lattice spacing. Assuming, that XZY is equilibrated, i.e., $\frac{d\rho_{XZY}}{dt} = 0$, at a much faster rate than any other reaction, we can calculate an effective rate of X production in the reaction, $XZ + Y \xrightarrow{k_2'} XXY$, as $k_2' = \frac{d^+ k_2}{d^- + k_2}$. Now we can immediately see that, when, $d^- < k_2$, $k_2'$ increases as $D_Y$ increases, and, that it saturates to a value $\pi k_2 / l_0^2$, at large $D_Y$ (when, $d^- \gg k_2$). The above analysis provides a qualitative understanding of the trends, but in order to obtain the quantitative change in velocity of the interface due to particle correlations and stochastic fluctuations one approach is to examine the Master equation for the system[27]. One can write down an action functional from the Master equation in terms of coherent states as a way to capture the effects of fluctuations[28,29] (details in the appendix). A naïve dimensional analysis of the terms in the action functional shows that the critical dimensions of the terms proportional to $k_1$ and $k_2$ are, $d_c = 2$ and $d_c = 1$, respectively. Thus, when $k_1 = 0$, one may naively assume that the fluctuations will not play any role in the system at d=2> $d_c = 1$. However, the effective field theory describing the interface motion can be very different than the one governing bulk dynamics, and fluctuations can still be important for the former[30]. Our simulations show that fluctuations indeed affect the interface motion, and we leave that analysis for a future work.

The effect of stochastic fluctuations is also manifested in the spreading of the width, $w(t)$, of the interface as time increases. The variation of $w(t)$ with time for different values of $L_y$ is shown in Fig.6, and the data can be collapsed on a master curve $w^2(t) = L^{2\alpha} f(t/L_y^z)$, with $\alpha = \beta z$. Asymptotically, $f(x) \sim x^{2\beta}$ when, $x \ll 1$ and $f(x) \sim const.$ when $x \gg 1$. These fits suggest that the roughness of the interface follows the scaling of the KPZ equation[31], where the exact scaling exponents in d=1, take the values, $\alpha = 1/2$, $\beta = 1/3$ and $z = 3/2$. As expected from universality, the scaling exponents do not depend on the value of $D_Y$.

**Conclusions:**

Using a simple model, we have shown how positive feedback in the presence of slow diffusion can give rise to nucleation of clusters of activated molecules, which then spread in space at a rate much faster than diffusion. The minimal model aims to mimic positive feedback regulation of signaling modules, such as Ras activation by the enzyme SOS. However, the framework is quite general and can be used to describe a variety of cell signaling processes[11] that are subject to feedback regulation. We also study the effect of fluctuations on the motion of the reaction/diffusion interface, and find that they lead to a rough interface no longer described by the mean-field kinetics. Interestingly, the fluctuations arising from positive feedback are irrelevant for description of bulk behavior.

**Acknowledgment:**

This work was supported by the NIH Directors Pioneer Award and 1PO1.

**Figure Captions:**

Fig. 1 **Monte Carlo rules for the model.** The Monte Carlo moves are shown on a lattice in d=1. The grey, green, red and white circles denote particles Z, Y, X, and an empty site, respectively. This color scheme is followed in all figures. (a) Shows a diffusive move with diffusion constant D. (b) Shows a pair-wise reaction between Z and Y particles with a rate $k_1$. (c) The possible positive feedback reactions between X, Y and Z particles with a rate $k_2$. (d) Annihilation of Y particles with a rate $k_3$.

Fig. 2 **Domain growth in d=2**. Particle configurations from the simulation in a 256 x 256 lattice. At t=0, the system starts with 30768 uniform randomly distributed Z particles (concentration = 0.469), and 2000 Y particles (concentration = 0.030). The parameters $k_1^{MC} = 0.001$, $D^{MC} = 10$, and $k_3^{MC} = 0$, are used for all simulations; configurations with $k_3 \neq 0$ are shown in the EPAPS (Fig.S1). (a) Configurations showing domain growth of the X particles (red) among the Y (green) and Z (grey) particles for a strong the positive feedback ($k_2 = 10$). Each MC step (MCS) corresponds to $\Delta t \approx 0.05$ s. (b) Configurations when the positive feedback reaction is turned off ($k_2 = 0$). Each MC step (MCS) corresponds to $\Delta t \approx 0.1$ s.

Fig. 3 **Characterization of nucleation and domain growth**. The structure factor S(q,t) is used to characterize the nucleation and domain growth in the system. All the data are averaged over 1000 initial configurations. (a) Shows S(q,t) vs q calculated at times, t = 800 MCS (circle), t = 1600 MCS (square), t = 2400 MCS (diamond) and t = 3200 MCS (triangle). The parameters are the same as in Fig.2a. (b) The data points for S(q,t) in part (a) are shifted by $\Delta(t)$, where $S(q \to \infty, t) + \Delta(t) = \langle \rho_X^2(t \to \infty) \rangle \approx 0.25$. For this choice $\Delta(t \to \infty) \to 0$, because at long times all the Z particles are converted into X particles which are eventually homogenously distributed in space; thus $S(q \neq 0, t \to \infty) \to \langle \rho_X^2(t \to \infty) \rangle$. (c) Shows S(q,t)+$\Delta$(t) for $D^{MC}$=100 at t = 2750 MCS (circle), t = 3850 MCS (square), t = 4950 MCS (diamond) and t = 6050 MCS (triangle). All other parameters are identical to that of Fig.3b. (d) Dependence of the wave-vector associated with the size of the critical nucleus, $q_c$, with the diffusion constant D. We determined the values of $q_c$ from the data points the where any two curves are, for the first time, separated by a distance that lies within the errorbars $\Delta_S$ in the S(q,t)(+offset\Delta) data in that region. The upper and lower limits in $q_c$ are calculated using, $\Delta_S = \Delta_S / 2$ and, $\Delta_S = \Delta_S$ respectively. The solid line shows a function of the form, $q_c = A D^{1/2}$, consistent with Eq.(2), for comparison.

Fig. 4 **Time dependence of production of X particles.** Plots of variations of the fraction $f_X(t)$ with time t, for different diffusion constants $D^{MC}$. The other parameters are identical to that of Fig.2. The solid lines are fits to the data with a function, $f_X(t) = 1 - Ae^{-B(t-t_0)^{2a}}$, where, A, B and $t_0$, are fitting parameters, we set $a$=1, for all the fits, corresponding to a linear growth in the size of domains of X particles with time.

Fig. 5 **Calculation of interface velocity from simulations.** (a) Variation of the mean interface position $\bar{h}(t)$ with time t, shown for different sizes of the interface. The box size along the direction of front propagation is set to $L_x = 512$ for all the simulations. The data are averaged over 1000 initial

configurations, and the values of the rate constants are, $k_1^{MC}=0$, $k_2^{MC}=10$ and $k_3^{MC}=0$; the diffusion constants are set to $D_X^{MC}=D_Y^{MC}=1$. (b) Shows the variation of the front velocity $v$, with the diffusion constant $D_Y^{MC}$. We have set $D_X^{MC}=1$ for all the simulations, and the other parameters are identical to that of Fig.5a. The system size for this calculation is 512x128. The initial concentrations of X+Z particles and Y particles are set to 0.5 and 0.03 respectively, and the data are again averaged over 1000 initial configurations. The solid dashed line shows the mean-field velocity predicted by the Fisher equation in Eq.(5).

Fig. 6 **Scaling behavior of the interface width**. The scaling of the width $w(t)$ of the interface is shown, at values of the simulation parameters identical to that of Fig.6a. The values of the scaling exponents that enable data collapse are, $z=3/2$, $\alpha=1$ and $\beta=1/3$, which are the exact values of the exponents of the KPZ equation in $d=1$.

**Appendix**

**Calculation of the Rate constants for the Ras-SOS system**

Activation of Ras through SOS can be described by the following reactions:

R1. $S+R_D \xleftrightarrow{(p_1,p_{-1})} SR_D \xrightarrow{p_{1f}} S+R_T$
R2. $S_a+R_T \xleftrightarrow{(p_2,p_{-2})} S_aR_T$
R3. $S_aR_T+R_D \xleftrightarrow{(p_3,p_{-3})} (S_aR_T)R_D \xrightarrow{p_{3f}} S_aR_T+R_T$

The short forms of the complexes used above are as indicated below:

$S \equiv SOS, R_D \equiv Ras-GDP, R_T \equiv Ras-GTP, SR_D \equiv SOS-Ras-GDP$,
$S_a \equiv$ allosteric site of $SOS, S_aR_T \equiv S_a-Ras-GTP, (S_aR_T)R_D \equiv (S_a-Ras-GTP)Ras-GDP$
$R_{Tot}=R_T+R_D+SR_D+S_aR_T, S_{Tot}=S+SR_D+S_aR_T$. The parameters $p_n$ and $p_{-n}$ indicate the binding and unbinding rates of the $n$th reaction respectively; $p_{nf}$

denotes the rate of a catalytic step. The ratios of the rates, $p_{1D} = p_{-1}/p_1$, $p_{2D} = p_{-2}/p_2$, $p_{3D} = p_{-3}/p_3$, and the catalytic rates $p_{1f}$, $p_{2f}$ have been measured in *in-vitro* experiments, and corresponding values are listed in Table 1.

Table 1: Values of the measured parameters in *in-vitro* experiments

| parameters | $p_1$ ($\mu m$)$^2$s$^{-1}$ molecules$^{-1}$ | $p_{-1}$ s$^{-1}$ | $p_{1f}$ s$^{-1}$ | $p_2$ ($\mu m$)$^2$s$^{-1}$ molecules$^{-1}$ | $p_{-2}$ s$^{-1}$ | $p_3$ ($\mu m$)$^2$s$^{-1}$ molecules$^{-1}$ | $p_{-3}$ s$^{-1}$ | $p_{3f}$ s$^{-1}$ |
|---|---|---|---|---|---|---|---|---|
| values | 0.0658[14] | 1.0[14] | 0.0005[14] | 0.1035[14] | 0.40 | 0.047[14] | 0.1[14] | 0.038[13] |

The above values are reported for the reactions occurring in d=2 in the plasma membrane; the binding rates in d=3 measured in experiments are converted to d=2 using, $(p_n)_{2D} = (p_n)_{3D}/\Delta$, where, $\Delta = 1.7nm$, which is the radius of gyration for a Ras molecule. All the reactions reported above take place on the plasma membrane; thus, the values of the rates at d=2 represent reaction kinetics in the physical situation.

When the concentration of the enzymes, $S$ and $S_a R_T$, are much smaller than that of the substrate, $R_D$, we can have $(p_{-1} + p_{1f})/p_1 + R_{Tot} \gg S_{Tot}$ and $(p_{-3} + p_{3f})/p_3 + R_{Tot} \gg [S_a R_T]$. In this situation, $d[SR_D]/dt \approx 0$ and $d[(S_a R_T)R_D]/dt \approx 0$ represent a very good approximation (Michaelis-Menten pseudo-state approximation[32]) to the dynamics. This approximation holds well in cells because the concentration of Ras (60 – 240 molecules/($\mu m$)$^3$) is larger than the concentration of SOS (20-60 molecules/($\mu m$)$^3$). In addition, SOS molecules can act on Ras molecules only when they are brought in the vicinity of the plasma membrane by Grb2 molecules associated with other signaling complexes. Thus, the concentration of SOS molecules that are responsible for activation of Ras can be even smaller than the measured values in *in-vitro* experiments. Furthermore, as time scales associated with $p_2$ and $p_{-2}$ are faster than that for the rates $p_{1f}$ and $p_{3f}$, $d[S_a R_T]/dt \approx 0$ is a good approximation at time-scales longer than that for catalysis. With the above approximations, we can write down the production of $R_T$ arising from reactions R1 (without positive feedback) and R2 (with positive feedback) as,

$$\frac{d[R_T]}{dt} = k_1 [S][R_D] + k_2 [S][R_T][R_D] \quad , \tag{A1}$$

where,

$$k_1 = \frac{p_{1f} p_1}{p_{-1} + p_{1f}} \quad , \tag{A2}$$

and

$$k_2 = \frac{p_{3f} p_3 p_2}{(p_{-3} + p_{3f}) p_{-2}}. \tag{A3}$$

Thus the values of the parameters $k_1$ and $k_2$ in Eq.(1) calculated for the Ras-SOS systems in $d=2$ are, $k_1 = 3.29 \times 10^{-5} (\mu m)^2 s^{-1}/(molecules)$ and $k_2 = 3.3 \times 10^{-3} (\mu m)^4 s^{-1}/(molecules)^2$.

**Derivation of the Mean-field equation from the Master Equation**

We derive the Master equation for the model described in the main text for $d=1$, but the procedure is easily generalized to higher dimensions. Then following a standard formalism for second quantization, we write down an action functional for the system. This action is then used to derive the mean field equations for the system.

The model is described by the set of variables, $\{n_i^X\}, \{n_i^Y\}$, and $\{n_i^Z\}$ which denote the occupancy (0 or 1) the lattice sites $\{i\}$. The probability of having a particular configuration evolves in time following a Master equation,

$$\frac{\partial P(\{n_i^X\},\{n_i^Y\},\{n_i^Z\}t)}{\partial t} = \text{terms arising from reactions} + \text{terms arising from diffusion}$$

The terms originating from the first reaction of Eq.(1) are,

$$R_1 = \sum_{i=1}^{N} k_1^{MC} [\delta_{n_i^X 1}\delta_{n_i^Y 0}\delta_{n_i^Z 0}\delta_{n_{i+1}^X 0}\delta_{n_{i+1}^Y 1}\delta_{n_{i+1}^Z 0} P(\{..n_i^X-1,..\},\{..,n_i^Y,..\},\{..n_i^Z+1,..\},t)$$
$$+ \delta_{n_i^X 0}\delta_{n_i^Y 1}\delta_{n_i^Z 0}\delta_{n_{i+1}^X 1}\delta_{n_{i+1}^Y 0}\delta_{n_{i+1}^Z 0} P(\{..n_{i+1}^X-1,..\},\{..,n_i^Y,..\},\{..n_{i+1}^Z+1,..\},t)$$
$$- (\delta_{n_i^X 0}\delta_{n_i^Y 0}\delta_{n_i^Z 1}\delta_{n_{i+1}^X 0}\delta_{n_{i+1}^Y 1}\delta_{n_{i+1}^Z 0} + \delta_{n_i^X 0}\delta_{n_i^Y 1}\delta_{n_i^Z 0}\delta_{n_{i+1}^X 0}\delta_{n_{i+1}^Y 0}\delta_{n_{i+1}^Z 1}) P(\{n_i^X\},\{n_i^Y\},\{n_i^Z\},t)]$$

(A4)

The first and second terms in the above expression represent the following reaction steps,

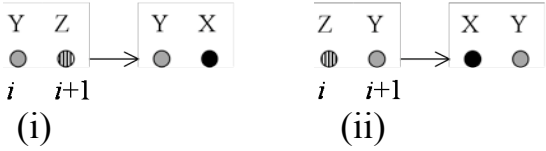

(i)   (ii)

The terms that will arise from the second reaction of Eq.(1) are,

$$R_2 = k_2^{MC} [\delta_{n_{i-1}^X 1}\delta_{n_{i-1}^Y 0}\delta_{n_{i-1}^Z 0}\delta_{n_i^X 1}\delta_{n_i^Y 0}\delta_{n_i^Z 0}\delta_{n_{i+1}^X 0}\delta_{n_{i+1}^Y 1}\delta_{n_{i+1}^Z 0} P(\{..n_i^X-1,..\},\{n_i^Y\},\{..n_i^Z+1,..\},t)$$
$$- \delta_{n_{i-1}^X 1}\delta_{n_{i-1}^Y 0}\delta_{n_{i-1}^Z 0}\delta_{n_i^X 0}\delta_{n_i^Y 0}\delta_{n_i^Z 1}\delta_{n_{i+1}^X 0}\delta_{n_{i+1}^Y 1}\delta_{n_{i+1}^Z 0} P(\{n_i^X\},\{n_i^Y\},\{n_i^Z\},t)$$
$$+ (5\times 2 \text{ similar terms for steps (iv) - (viii))}]$$

(A4)

The terms contributing to the above expression originate from the following processes:

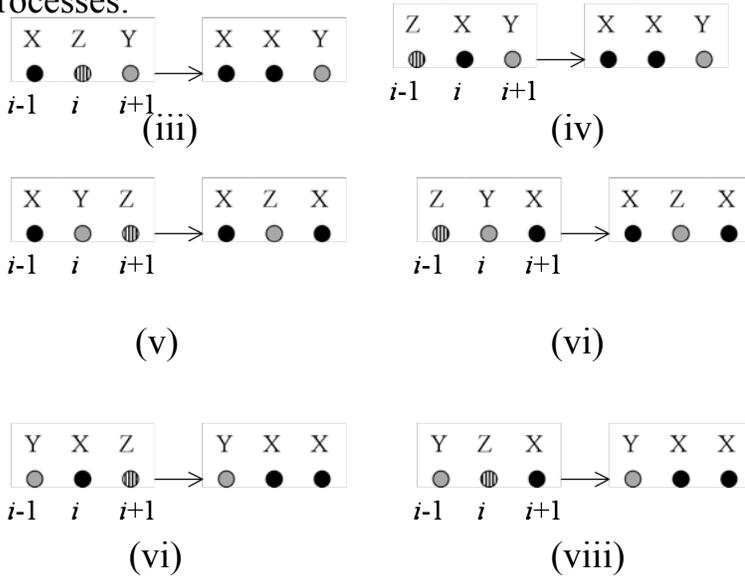

(iii)   (iv)

(v)   (vi)

(vi)   (viii)

The decay of the Y particle arising from the third reaction in Eq.(1) is represented by the following terms,

$$R_3 = \sum_{i=1}^{N} k_3^{MC} [\delta_{n_i^X 0} \delta_{n_i^Y 0} \delta_{n_i^Z 0} P(\{..n_i^X,...\},\{..,n_i^Y+1,..\},\{..n_i^Z,...\},t)$$
$$-\delta_{n_i^X 0} \delta_{n_i^Y 1} \delta_{n_i^Z 0} P(\{n_i^X\},\{n_i^Y\},\{n_i^Z\},t)] \tag{A5}$$

The above process corresponds to,

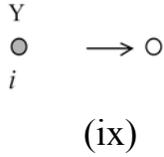

(ix)

The diffusion moves will introduce the following terms,

$$D^{MC}[\delta_{n_i^X 0} \delta_{n_i^Y 0} \delta_{n_i^Z 0} \delta_{n_{i+1}^X 1} \delta_{n_i^Y 0} \delta_{n_i^Z 0}[P(\{...,n_i^X+1,n_{i+1}^X-1,...\},\{n_i^Y\},\{n_i^Z\},t)-P(\{n_i^X\},\{n_i^Y\},\{n_i^Z\},t)$$
+ similar terms for diffusion of Y and Z particles]

(A6)

The above terms correspond to the particle hops depicted below,

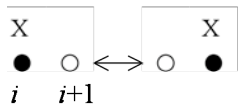 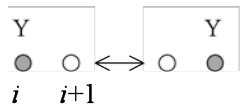 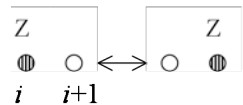

(x)　　　　　　　　(xi)　　　　　　　　(xii)

The master equation can be cast in a second quantized form,

$\frac{\partial |\psi\rangle}{\partial t} = -\hat{H}|\psi\rangle$, where $|\psi\rangle = \sum_{\{n_i^X\}\{n_i^Y\}\{n_i^Z\}} P(\{n_i^X\},\{n_i^Y\},\{n_i^Z\},t)|\{n_i\}\rangle$, and the Hamiltonian

operator $\hat{H}$, is a function of the bosonic creation and annihilation operators $\{\bar{a}_i^X, a_i^X\}, \{\bar{a}_i^Y, a_i^Y\}$ and $\{\bar{a}_i^Z, a_i^Z\}$ for the X, Y and Z particles. An action functional S, can be constructed from the Hamiltonian in terms of the coherent states, $\{\hat{\phi}_i^X, \phi_i^X\}, \{\hat{\phi}_i^Y, \phi_i^Y\}$, and $\{\hat{\phi}_i^Z, \phi_i^Z\}$. The process of constructing the action functional is standard and can be found in several references[28,33].

Next we take the continuum limit, by performing the following operations:
$$\phi_i(t) \to l_0 \phi(x,t); \hat{\phi}_i(t) \to \hat{\phi}(x,t); \sum_i \to l_0^{-1} \int dx;$$

$$\phi_i(t) \to l_0(\phi(x,t) + l_0 \partial_x \phi(x,t) + l_0^2/2 \partial_x^2 \phi(x,t) + \text{higher order terms}$$
$$\hat{\phi}_i(t) \to l_0(\hat{\phi}(x,t) + l_0 \partial_x \hat{\phi}(x,t) + l_0^2/2 \partial_x^2 \hat{\phi}(x,t) + \text{higher order terms}$$

where $l_0$ is the lattice spacing. This can be easily generalized to $d$ dimensions where we get the following action:

$$S = \int d^d x \int_0^t dt [\hat{\phi}^a \partial_t \phi^a - D^{MC} l_0^2 \hat{\phi}^a \nabla^2 \phi^a e^{-2v_0 \hat{\phi}^a \phi^a} + 2k_1^{MC} dv_0 (\hat{\phi}^Z \phi^Z - \hat{\phi}^X \phi^Z)(\hat{\phi}^Y \phi^Y) e^{-2v_0 \hat{\phi}^a \phi^a}$$
$$+ 6k_2^{MC} dv_0^2 (\hat{\phi}^Z \phi^Z - \hat{\phi}^X \phi^Z)(\hat{\phi}^Y \phi^Y)(\hat{\phi}^X \phi^X) e^{-3v_0 \hat{\phi}^a \phi^a} + k_3^{MC} (\hat{\phi}^Y - 1) \phi^Y e^{-v_0 \hat{\phi}^a \phi^a}]$$
(A7)

In the above expression $v_0 = l_0^d$, and we have used the summation convention, $\hat{A}^a A^a = \hat{A}^X A^X + \hat{A}^Y A^Y + \hat{A}^Z A^Z$. The exponential terms in the action arise because of the hardcore repulsion between the particles. We expand such terms to linear order as $e^{-v_0 \hat{\phi}^a \phi^a} \approx 1 - v_0 \hat{\phi}^a \phi^a$, since the higher order terms in the expansion gives rise to dimensionally irrelevant contributions (with critical dimensions $d_c < 1$). Dimensional analysis indeed reveals that various nonlinear terms have critical dimension of $d_c = 2$, or $d_c \leq 1$. We find that all the terms proportional to $k_2$ have $d_c \leq 1$. The mean-field equations can be derived by extremizing the action with respect to $\hat{\phi}_i^X, \phi_i^X, \hat{\phi}_i^Y, \phi_i^Y$, and $\hat{\phi}_i^Z, \phi_i^Z$. Derivatives of the action with respect to $\hat{\phi}^a$ ($a = \{X, Y, Z\}$) are zero for $\hat{\phi}^a = 1$. The remaining equations in $\phi^a$ are given by,

$$\frac{\partial \phi^X}{\partial t} = -\frac{\delta H}{\delta \hat{\phi}^X} = (D^{MC} l_0^2 \nabla^2 \phi^X + 2k_1^{MC} dv_0 \phi^Z \phi^Y) e^{-2v_0 \sum_a \phi^a} + 6k_2^{MC} dv_0^2 \phi^X \phi^Z \phi^Y e^{-3v_0 \sum_a \phi^a}$$

$$\frac{\partial \phi^Z}{\partial t} = -\frac{\delta H}{\delta \hat{\phi}^Z} = (D^{MC} l_0^2 \nabla^2 \phi^Z - 2k_1^{MC} dv_0 \phi^Z \phi^Y) e^{-2v_0 \sum_a \phi^a} - 6k_2^{MC} dv_0^2 \phi^X \phi^Z \phi^Y e^{-3v_0 \sum_a \phi^a}$$

(A8)

$$\frac{\partial \phi^Y}{\partial t} = -\frac{\delta H}{\delta \hat{\phi}^Y} = D^{MC} l_0^2 \nabla^2 \phi^Y e^{-2v_0 \sum_a \phi^a} - k_3^{MC} \phi^Y e^{-v_0 \sum_a \phi^a}$$

(A9)

(A10)

Though, the fields $\phi^a$ are not particle densities, one can relate them to the actual densities by the transformations $\phi^a = \rho^a e^{-\hat{\rho}^a}$ and $\hat{\phi}^a = e^{\hat{\rho}^a}$. In the mean-field approximation, $\hat{\rho}^a = 0$ and $\rho^a = \phi^a$. Therefore, we get the following reaction diffusion equations for the density fields,

$$\frac{\partial \rho^X}{\partial t} = (D^{MC} l_0^2 \nabla^2 \rho^X + 2k_1^{MC} dv_0 \rho^Z \rho^Y) + 6k_2^{MC} dv_0^2 \rho^X \rho^Z \rho^Y$$
$$- (D^{MC} l_0^2 \nabla^2 \rho^X + 2k_1^{MC} dv_0 \rho^Z \rho^Y)(2v_0 \sum_a \rho^a \rho^a) + O((\rho^a)^4)$$

(A11)

$$\frac{\partial \rho^Z}{\partial t} = (D^{MC} l_0^2 \nabla^2 \rho^Z - 2k_1^{MC} v_0 \rho^Z \rho^Y) - 6k_2^{MC} dv_0^2 \rho^X \rho^Z \rho^Y$$
$$- (D^{MC} l_0^2 \nabla^2 \rho^Z - 2k_1^{MC} dv_0 \rho^Z \rho^Y)(2v_0 \sum_a \rho^a \rho^a) + O((\rho^a)^4)$$

(A12)

$$\frac{\partial \rho^Y}{\partial t} = D^{MC} l_0^2 \nabla^2 \rho^Y - k_3^{MC} \rho^Y - (D^{MC} l_0^2 \nabla^2 \rho^Y - k_3^{MC} \rho^Y) 2v_0 \sum_a \rho^a \rho^a$$

(A13)

The last and the higher order terms arise in the equations because of the site restriction constraints.

Naïve scaling analysis:

Scaling dimensions of the parameters in the above action are as follows:

The fields scale as, $[\phi^a] = L^{-d}$; $[\hat{\phi}^a] = L^0$; where, L is a length scale, therefore, the rate constants scale as,

$[k_1^{MC} v_0] = L^{d-2}$; $[k_2^{MC} v_0^2] = L^{d-1}$; $[k_3^{MC}] = L^{-z}$; $[D^{MC}] = L^{-z}$, with $z = 2$. As from the naïve scaling of the rate constants, the critical dimension ($d_c$) for the positive feedback term is $d_c = 1$ and that for the production of X particles by the first reaction in Eq.(1) is $d_c = 2$. By expanding the exponentials in the action one can see that terms higher than the second order give rise to irrelevant terms. Therefore, we can expect that when diffusion is important, the mean field dynamics for the density fields will hold for the positive feedback, and production of X species, for dimensions d >1, and d >2, respectively. However, the fluctuations in the interfacial kinetics can still be relevant in dimensions higher than the critical dimensions. This seems to be the case for our system, where in d=2, the interactions originating from the positive feedback gives rise to rough interfaces.

**Connection between the MC and the physical parameters**

Comparing the reaction diffusion equations that would correspond to Eq.(1) with Eqns.A11-A13 shows, $k_1 = 2k_1^{MC} dl_0^d$, $k_2 = 6k_2^{MC} dl_0^{2d}$, $k_2 = k_3^{MC}$, and $D = D^{MC} l_0^2$. The length and time scales in the simulation are chosen by setting $l_0 = 1$, and $D^{MC} = 10$, in the simulation to a length scale of $0.1 \mu m$ and a diffusion constant of $D = 0.1 (\mu m)^2 / s$. With this choice $k_1^{MC} = 0.0001$ and $k_2^{MC} = 10$ correspond to $k_1 = 4.0 \times 10^{-5} (\mu m)^2 s^{-1} / molecules$ and $k_2 = 12.0 \times 10^{-3} (\mu m)^4 s^{-1} / (molecules)^2$.

Fig.1

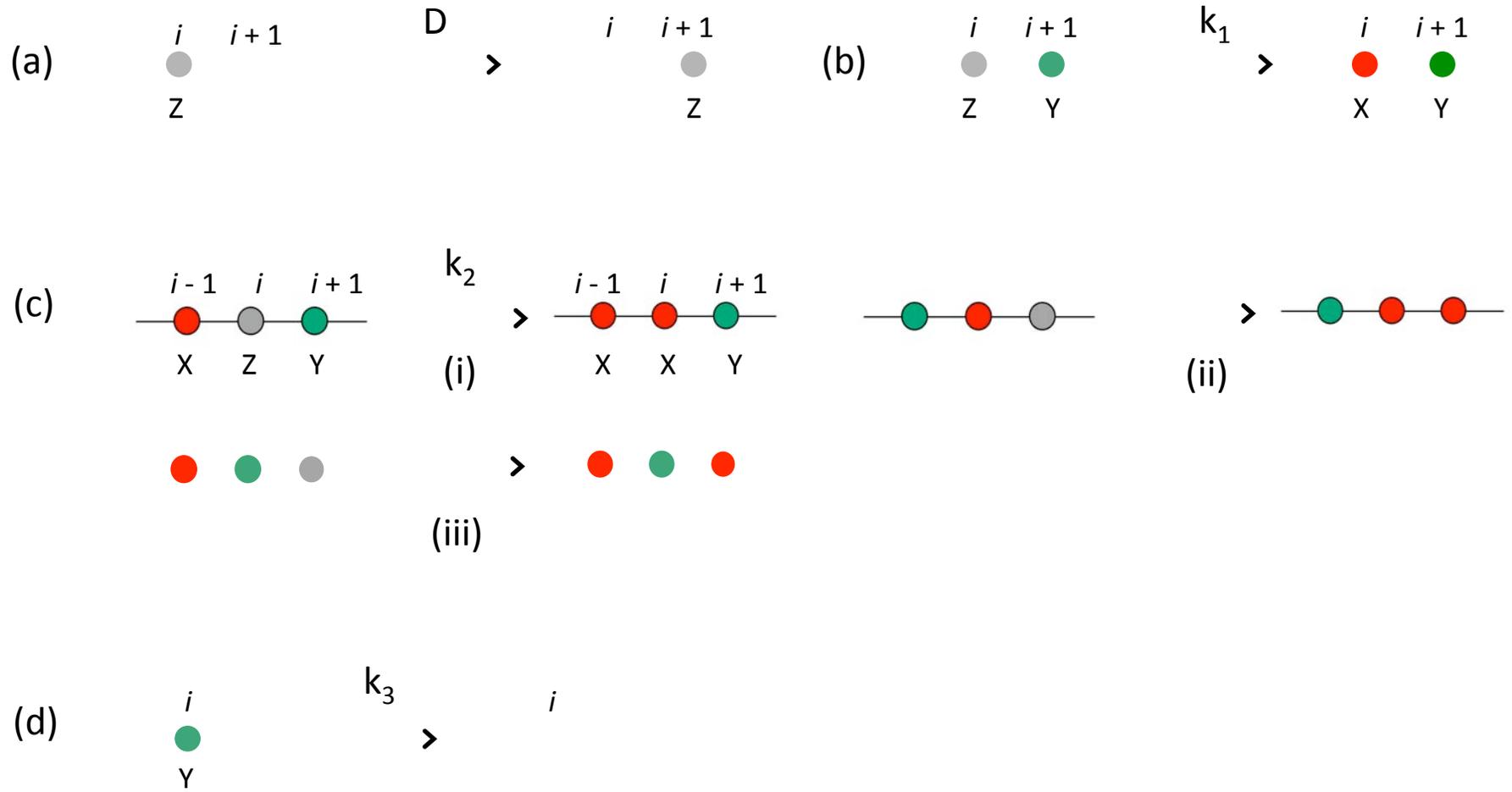

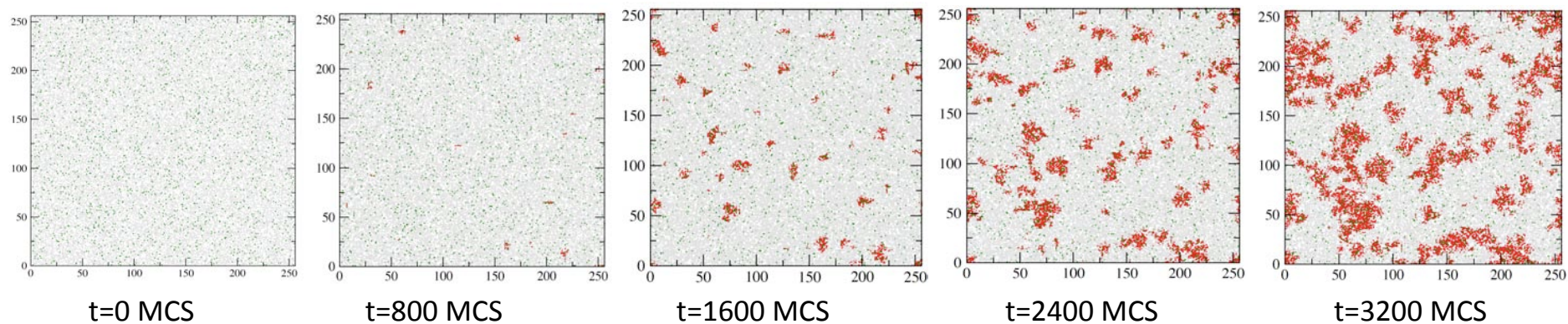

(a)

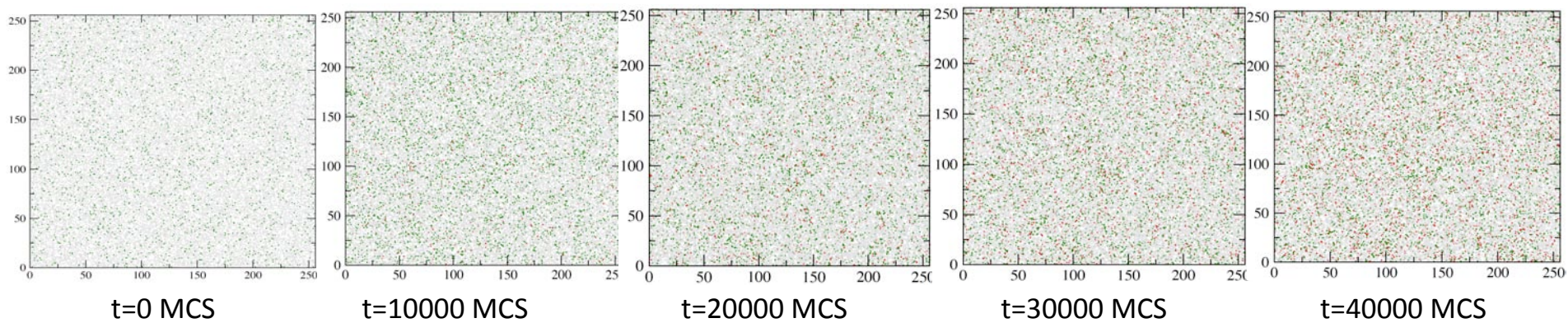

(b)

Fig. 2

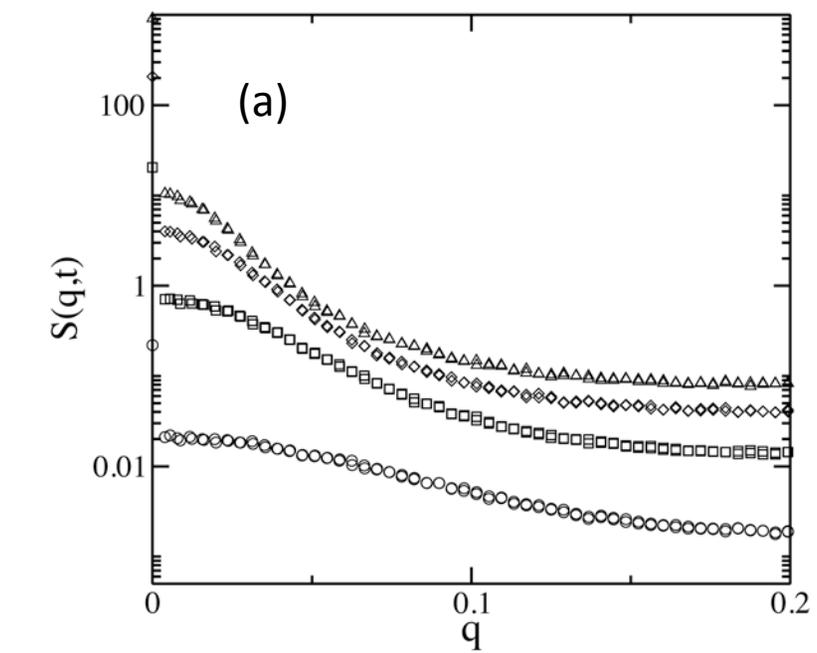 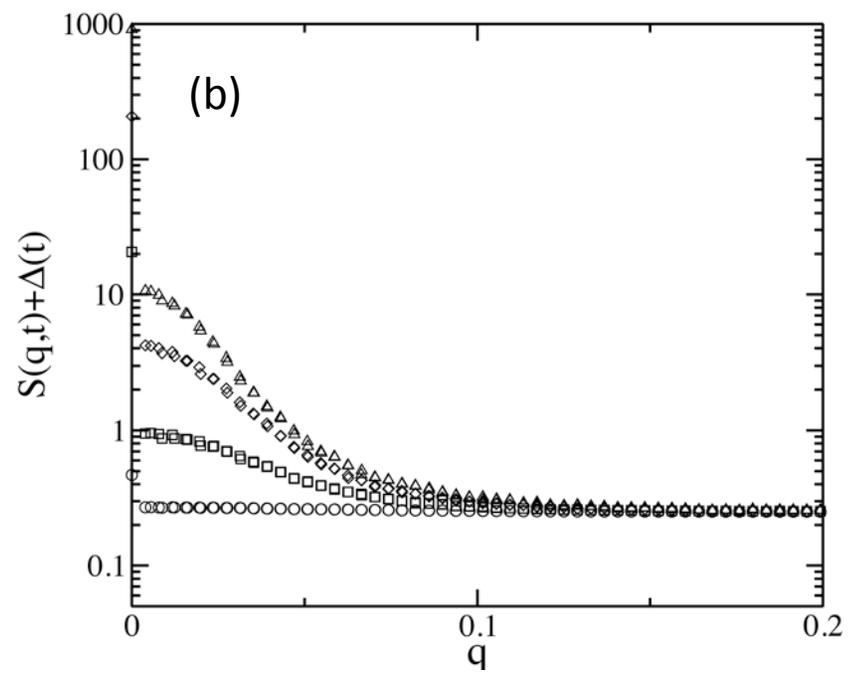
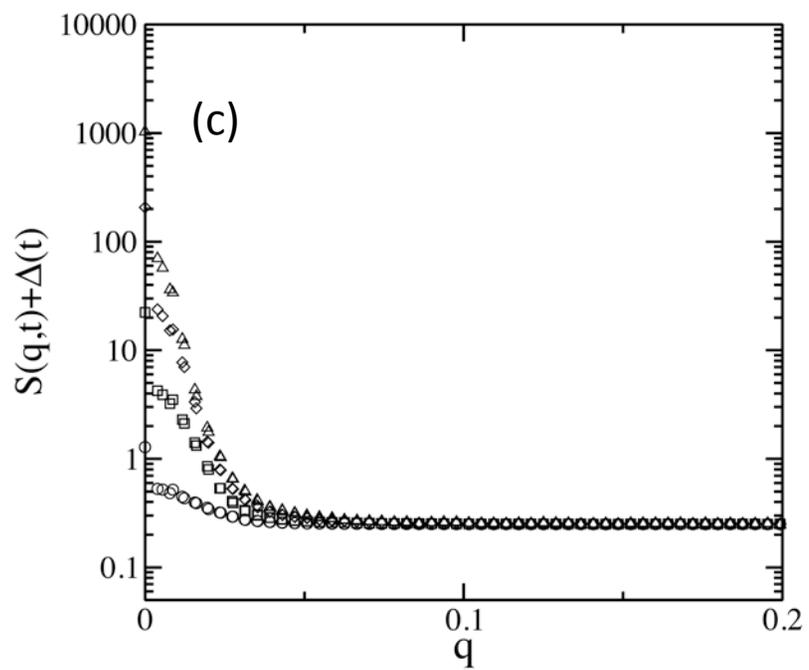 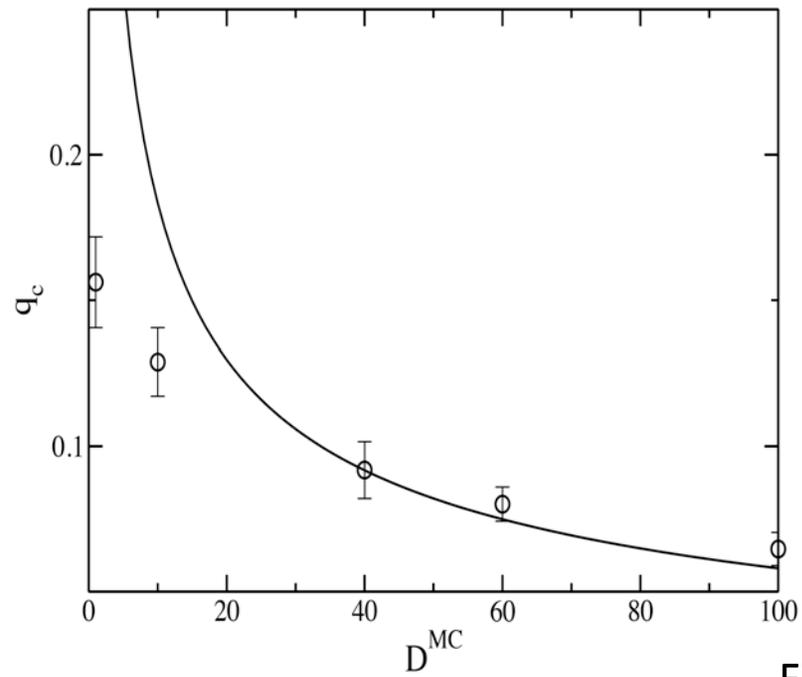

Fig 3

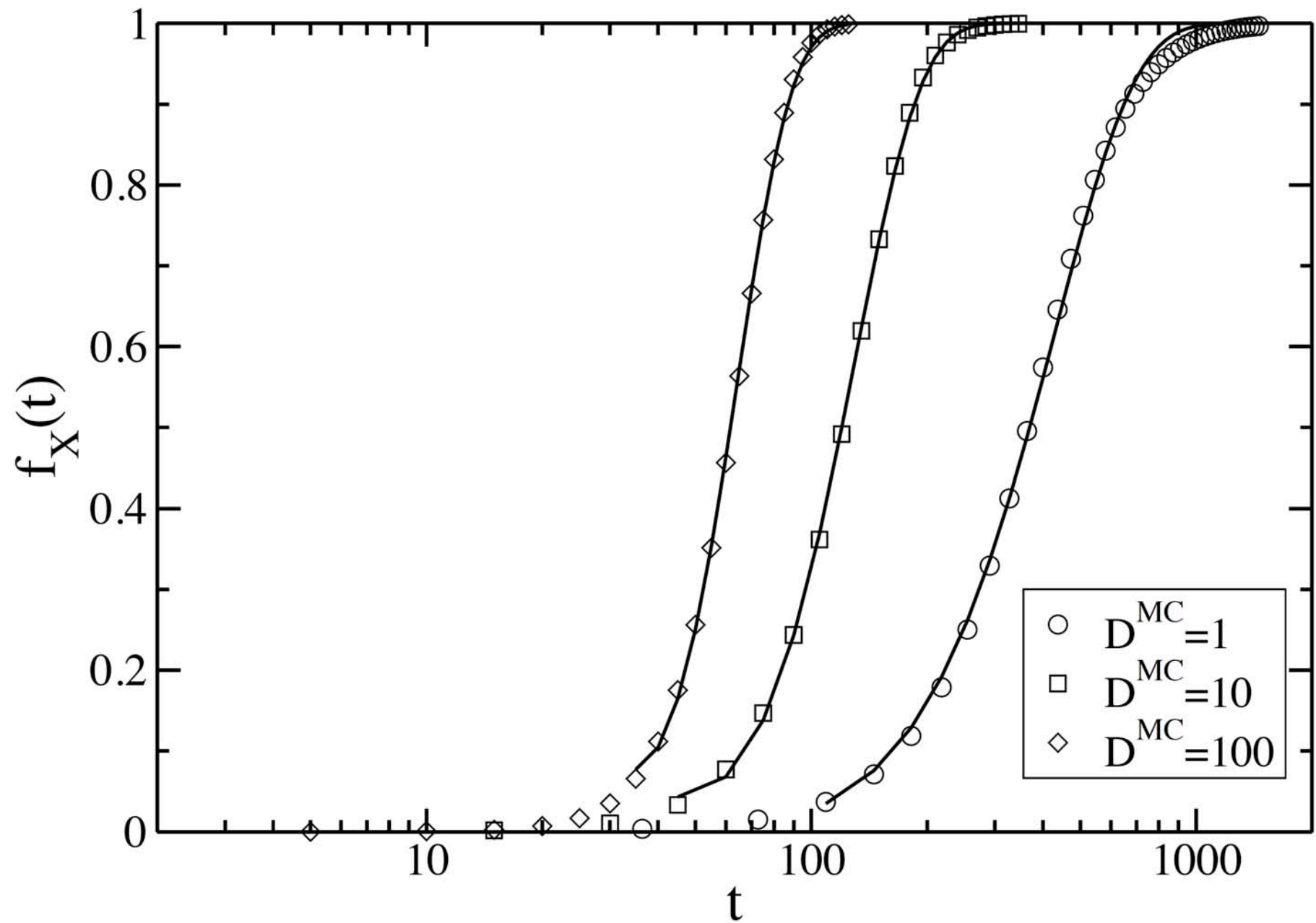

Fig. 4

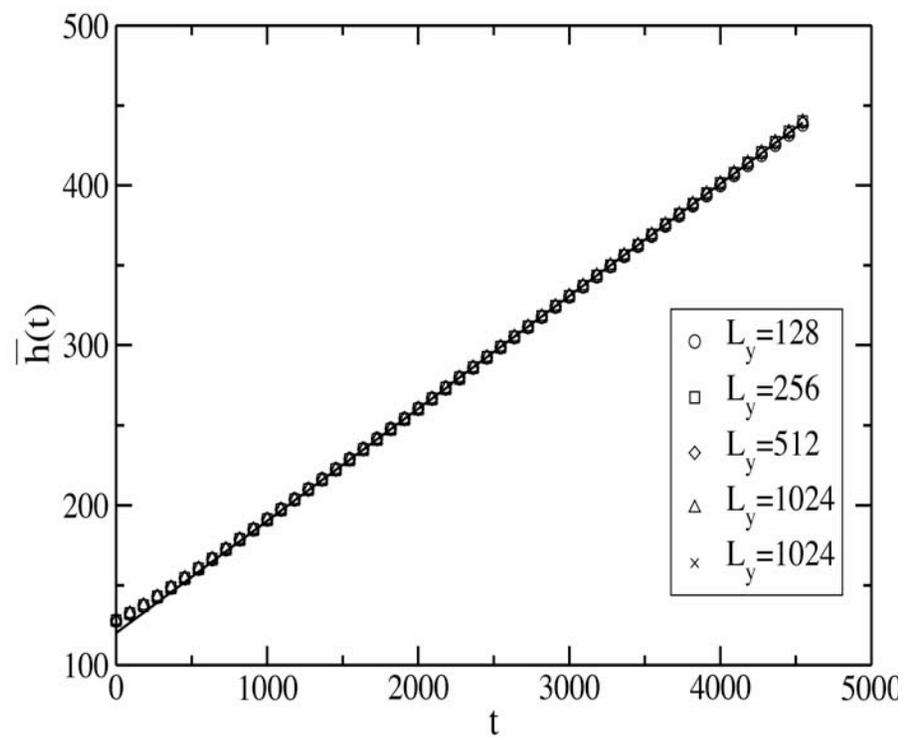 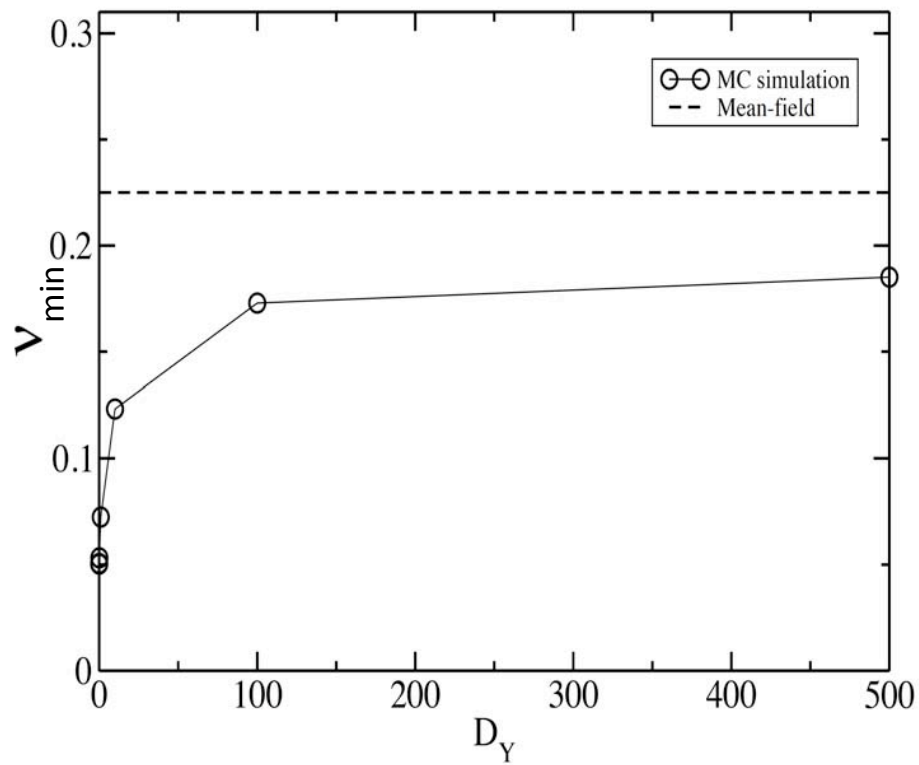

(a) (b)

Fig. 5

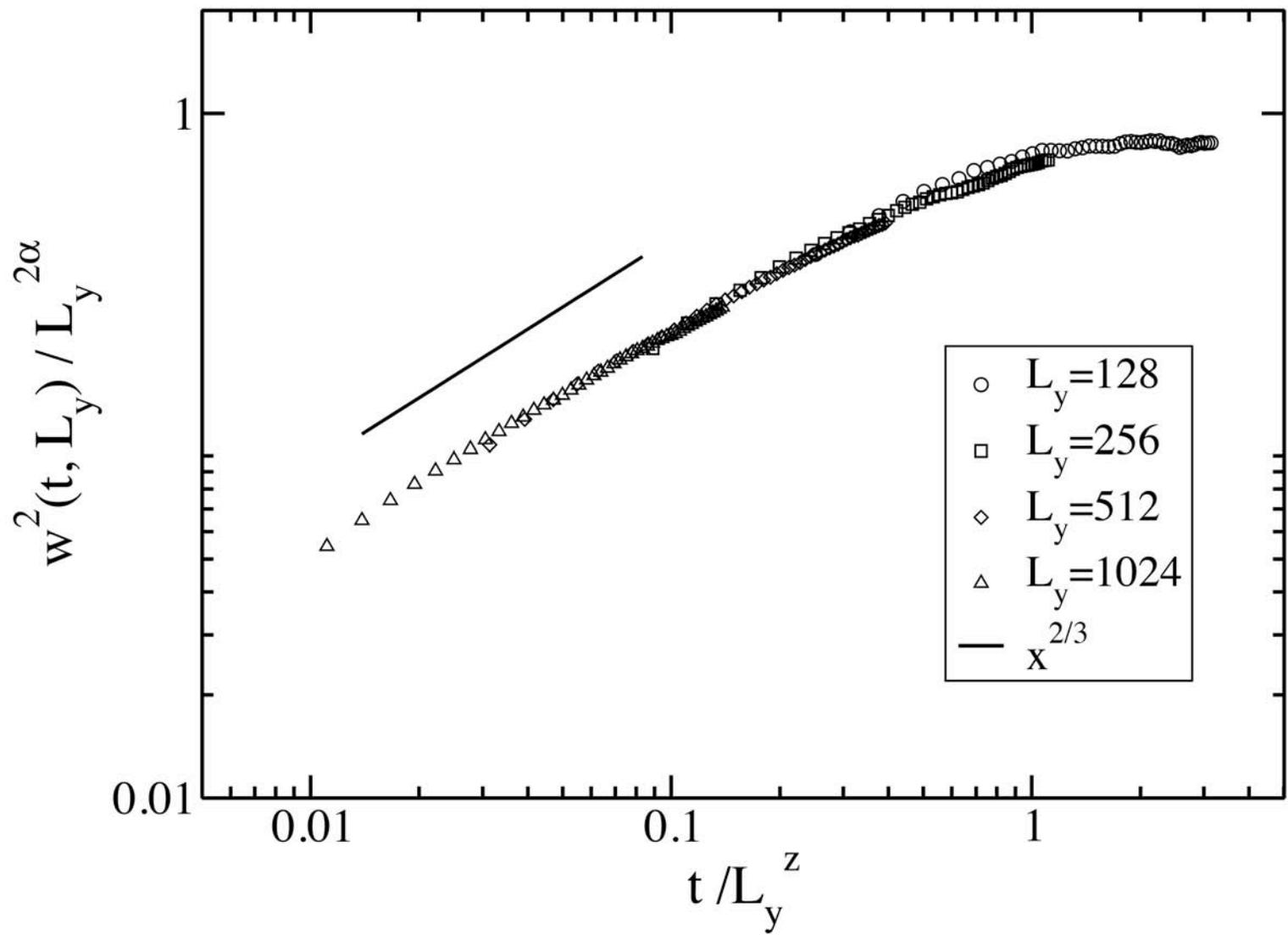

Fig. 6

EPAPS for the paper "Positive Feedback Regulation Results in Spatial Clustering and Fast Spreading of Active Signaling Molecules on a Cell Membrane"

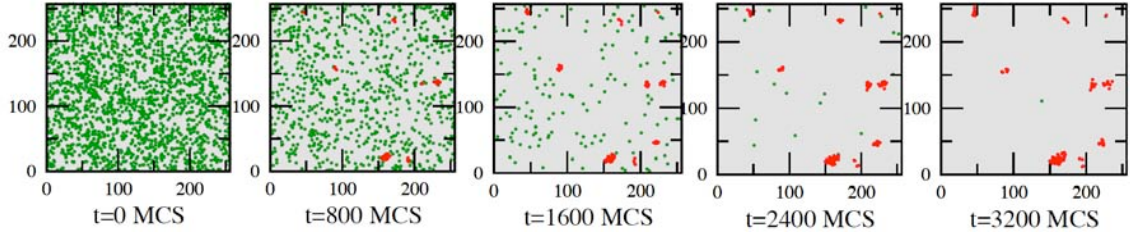

Fig. S1 **Effect of non-zero $k_3$ on domain growth**. The particle configurations are shown for $k_3$=0.1. All other parameters are the same as in Fig. 2.

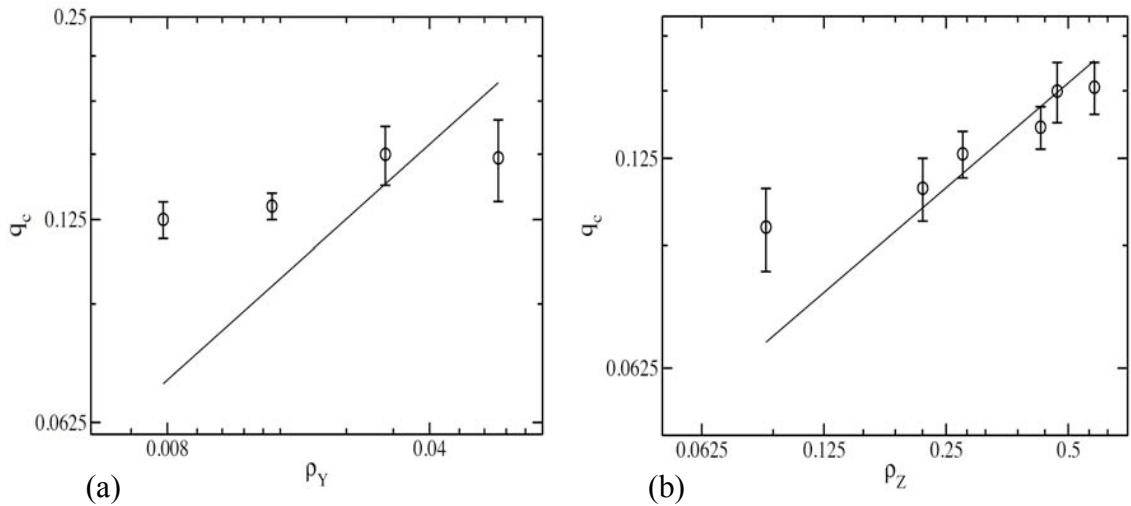

Fig. S2 **Variation of the critical nucleus size ($q_c$) with the concentrations of Y ($\rho_Y$) and Z ($\rho_Z$) particles.** (a) Shows the variation of $q_c$ in the log scale as $\rho_Y$ is increased. (b) Shows the variation as $\rho_Z$ is increased. All the other parameters are the same as in Fig. 2. The solid lines are fits to data to a function, $f(x) = ax^{1/2}$ (*a* is a fitting parameter) showing the variation of $q_c$ as predicted by Eq. 2.